\documentclass{ws-procs9x6}

\begin{document}

\title{ASTROPHYSICAL ORIGINS OF THE HIGHEST ENERGY COSMIC RAYS}

\author{SUSUMU INOUE}

\address{National Astronomical Observatory of Japan,\\
2-21-1 Osawa, Mitaka, Tokyo 181-8588, Japan\\
E-mail: inoue@th.nao.ac.jp}

\begin{abstract}
Theoretical aspects of potential astrophysical sources
of the highest energy cosmic rays are discussed,
including their energy budget and some issues on particle escape and propagation.
After briefly addressing AGN jets and GRBs,
we highlight the possibility of heavy nuclei originating from cluster accretion shocks.
The importance of X-ray and gamma-ray signatures in addition to neutrinos
as diagnostic tools for source identification is emphasized.
\end{abstract}

%\keywords{Style file; \LaTeX; Proceedings; World Scientific Publishing.}

\bodymatter

\section{Introduction}
\label{sec:intro}

Several decades after their discovery,
the origin of ultra-high energy cosmic rays (UHECRs),
cosmic particles with energies $10^{18}$-$10^{20}$ eV and above,
remains one of the biggest mysteries in physics and astrophysics \cite{nw00,hil06}.
Many issues contribute to the difficulty of the problem.
For experimentalists, the extremely low event rates
necessitate detector facilities with huge effective area in order to obtain reliable results.
The relevant energies far exceed those of terrestrial experiments,
often making the determination of basic observables such as particle energy and composition
dependent on interaction models with large uncertainties.
On the theoretical side, conceiving viable explanations for the production of UHECRs
with conventional physical mechanisms in known astrophysical objects is a great challenge. 
The unavoidable yet uncertain influence of Galactic and extragalactic magnetic fields
on UHECR propagation pose further complications.

However, great advances are expected in the coming years
with the advent of new generation facilities
such as the Pierre Auger Observatory and the Telescope Array,
as well as future projects such as the Extreme Universe Space Observatory.
Combined with crucial complementary information from
neutrino, X-ray and gamma-ray observatories,
the solution of the mystery could be within sight soon.
This article, by no means a thorough review,
discusses selected theoretical topics in this exciting field,
focusing on the astrophysical aspects.

\section{Issues on UHECR propagation}
\label{sec:prop}

We first touch upon some issues concerning the propagation of UHECRs,
the basics of which have been well reviewed elsewhere (e.g. Ref. \refcite{bs00}).
The observed global isotropy in the arrival directions strongly suggests
that UHECRs are of extragalactic origin.
If UHECRs are protons, photopion interactions with cosmic microwave background (CMB) photons
must induce severe energy losses at $\gtrsim 7 \times 10^{19}$ eV
for propagation distances $\gtrsim 30$ Mpc.
Unless the sources lie much nearer,
a spectral (``GZK'' \cite{gzk66}) cutoff is expected above these energies.
Whether or not there is actual evidence for this in the observed UHECR spectrum
is a matter of controversy at the moment, and will not be discussed here.

Here, we call attention to the possibility that the highest energy UHECRs
are composed mainly of heavy nuclei such as iron.
For nuclei, the dominant energy loss process above $10^{19}$ eV during intergalactic propagation
is photodisintegration and pair production interactions with photons
of the far-infrared background (FIRB) and the CMB \cite{psb76}.
Evaluations based on recent determinations of the FIRB
show that the energy loss distance for iron nuclei at $10^{20}$ eV is $\gtrsim 100$ Mpc,
somewhat larger than that for protons \cite{ss99}.
Observationally, the nuclear composition of UHECRs,
particularly at the highest energies, is quite uncertain.
Important clues come from fluorescence measurements
of shower elongation rates by the HiRes experiment,
indicating that the composition changes from heavy-dominated below $10^{18}$ eV
(presumably of Galactic origin) to light-dominated above this energy \cite{abb05}.
However, above $\sim 3 \times 10^{19}$ eV
the statistics run out and there is no reliable information currently available.
Even for the lower energy range,
other methods give rather different results \cite{ave03},
and systematic uncertainties in interaction models and atmospheric optical attenuation
remain a serious concern \cite{wat04}.
In any case, at present, it is quite viable that UHECRs at $\gtrsim 3 \times 10^{19}$ eV
are predominantly heavy nuclei originating from extragalactic sources.
This picture will be elaborated on in Sec.\ref{sec:clus}.

Another crucial issue that is often underestimated
is deflection by extragalactic and Galactic magnetic fields,
which can affect the arrival directions of UHECRs
and also act to lengthen their effective propagation distance.
The strength and distribution of magnetic fields in the intergalactic medium
and at high latitudes in the Galaxy are very poorly known, both observationally and theoretically.
Faraday rotation measurements of distant radio sources
give only upper limits in the nanogauss range for intergalactic fields on average,
subject to assumptions on the field coherence scale and ionized gas distribution \cite{kro94}.
Realistically, whatever their origin, intergalactic fields can be expected
to have some correlation with the distribution of large scale structure.
Attempts to model this through cosmological simulations haven given different results
depending on the input physics and numerical methods \cite{sme03, dol04}.
The effect of Galactic magnetic fields is also model-dependent \cite{tak06}.
Definitive answers may not come until the operation of the Square Kilometer Array (SKA),
slated to undertake an ``all-sky'' survey of rotation measures
toward $>10^7$ background sources with typical angular separations $\sim 90''$,
out to redshifts  $z \gtrsim 3$ \cite{gae06}.
Until such information becomes available,
we cannot rule out the possibility that deflections by intervening magnetic fields
are significant even for the highest energy CRs.

\section{Candidate astrophysical sources of UHECRs}
\label{sec:source}

A minimum requirement for astrophysical sources of UHECRs
is the ability to magnetically confine particles of the requisite energies.
For particles with energy $E$ and charge $Z$, this implies the condition
$(R/{\rm pc}) (B/{\rm 1 G}) \gtrsim (E/{\rm 10^{20} eV})/Z$
between the system's size $R$ and magnetic field $B$.
Only a select few types of objects are known to meet this criterion,
among them the jets of radio-loud active galactic nuclei (AGNs),
gamma-ray bursts (GRBs), and clusters of galaxies \cite{hil84}.
Notwithstanding other candidates, in what follows, we focus on these three
as representative types of potential UHECR sources
(see Ref. \refcite{hil84} for other possibilities).
The actual maximum energy attainable
under different circumstances must be evaluated
by comparing the timescales for particle acceleration, usually
that for the first order Fermi mechanism in shocks,
against the timescales for limiting processes such as
source lifetime, particle escape, adiabatic or radiative energy loss, etc.

\begin{figure}
\begin{center}
\psfig{file=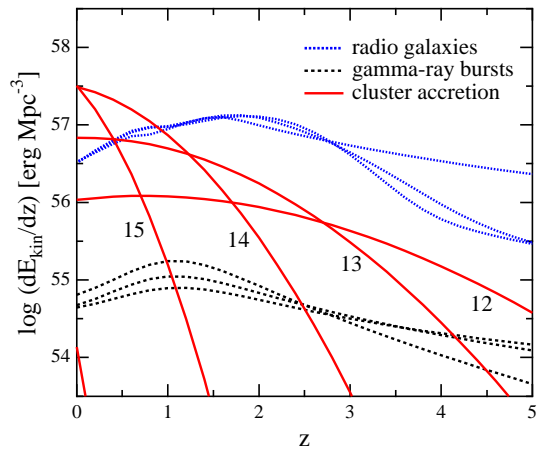,width=3.0in}
\end{center}
\caption{Energy budget of candidate UHECR sources:
AGN jets (dotted), GRBs (dashed) and cluster accretion shocks (solid),
the latter separately for each $\log M$ as labelled.}
\label{fig:ene}
\end{figure}

Equally important is the available energy budget.
Fig.\ref{fig:ene} shows estimates of the kinetic energy output averaged over the universe
as a function of redshift $z$ due to AGN jets, GRB explosions and accretion onto clusters,
which should be proportional to their cosmic ray output.
The plotted quantity is differential per unit $z$,
$dE_{\rm kin}/dz = (dt/dz) \int L (dn/dL) dL$,
where $L$ is the kinetic luminosity per object
and $dn/dL$ is the $z$-dependent luminosity function,
with cosmological parameters $h$=0.7, $\Omega_m$=0.3 and $\Omega_\Lambda$=0.7.
For AGN jets, we have made use of the observed radio luminosity function
along with the observed correlation between the radio and jet kinetic luminosities
of radio galaxies \cite{is01}.
GRBs were assumed to occur each with kinetic energy $E_{\rm GRB}=10^{53}$ erg
at a rate that follows the star formation history
and matches the $\log N$-$\log S$ distribution observed by BATSE \cite{pm01}
(note that this estimate is roughly independent of the beaming factor).
The three curves each for AGNs and GRBs correspond to
different evolutionary assumptions at the highest $z$,
with only small differences at low $z$.
Cluster accretion will be discussed in Sec.\ref{sec:clus}.

The results at low $z$ can be compared with the observed
energy density of UHECRs,
$\simeq 3 \times 10^{-20} {\rm erg \ cm^{-3}} \simeq 10^{54} {\rm erg \ Mpc^{-3}}$ above $10^{19}$ eV.
Considering further factors for energy loss during propagation,
it is apparent that whereas AGN jets and cluster accretion shocks
have reasonable margins to accommodate the energetics of UHECRs,
GRBs, with a substantially smaller energy budget,
require a very high efficiency of energy conversion into UHECRs,
a fact that has already been noted \cite{hil06,bgg06}.

For both AGN jets and GRBs, different locations along the outflow
can be UHECR production sites,
since one expects $B \propto R^{-1}$ under the naive assumption
that the ratio of magnetic to kinetic energy is constant.
One candidate in AGNs is the inner jet region
with $R \sim 10^{16}$-$10^{17}$ cm and $B \sim 0.1$-$1$ G,
known through observations of blazars
to be a site of particle acceleration, perhaps due to internal shocks.
Estimates show that the maximum proton energy should be limited
by photopion interactions with low frequency internal radiation
to somewhat below $10^{20}$ eV \cite{mpr01}.
Furthermore, efficient conversion to neutrons may be necessary
to allow the particles to escape the jet without suffering adiabatic expansion losses
and contribute to UHECRs by decaying back to protons outside.
A more promising site may be the hot spots of powerful radio galaxies,
termination shocks where large-scale jets are decelerated by the external medium,
with $R \sim 10^{21}$ cm and $B \sim 1$ mG.
Here the maximum energy may exceed $10^{20}$ eV, limited by escape \cite{bs87}.
However, note that these particles must further traverse
the extensive cocoon of shocked, magnetized jet material
in order to completely escape the system and constitute UHECRs,
an issue that has not been examined in detail.
A further possibility is
acceleration by the bow shocks
being driven into the ambient gas by the expansion of the cocoon \cite{nma95}.

In order to verify an AGN origin,
detailed analysis of observed UHECR arrival directions
and cross correlations with known source positions will undoubtedly be essential.
Nonetheless, given the uncertainties in intervening magnetic fields (Sec. \ref{sec:prop}),
it will also be highly desirable to have some means to pinpoint individual sources
through characteristic, UHECR-induced signatures of secondary neutral radiation.
Neutrinos are ideal in providing an unambiguous earmark of high energy hadrons \cite{hh02}.
However, even with km$^{3}$ detector facilities such as IceCube or KM3NeT,
it may not be easy to resolve individual AGNs \cite{ach05}.
Thus, distinctive electromagnetic signals will also be extremely valuable.
For radio galaxy hot spots,
synchrotron emission from UHE protons can produce nonthermal X-rays
that could be distinguished from other processes such as electron inverse Compton
through multiwavelength observations \cite{aha02}.
If the medium surrounding the source is sufficiently magnetized,
UHE protons propagating
in the source's vicinity can lead to diffuse gamma-ray emission from photomeson-triggered cascades
that may be detectable by current and upcoming instruments \cite{ga05}.

Potential locales for UHECR acceleration in GRBs
include internal shocks, external reverse shocks, and external forward shocks,
believed to be the emission sites of the prompt X-rays and gamma-rays,
optical flash and radio flare, and the radio to X-ray afterglow, respectively \cite{wax95}.
The external forward shock could be disfavored due to its ultrarelativistic velocity,
but this is controversial \cite{vdg03}.
For the mildly relativistic internal and external reverse shocks,
a different problem is that for the particles to escape the acceleration site without significant losses,
neutron conversion may be required,
as with AGN inner jet regions \cite{asa05}.
This entails some reduction in efficiency, which cannot be too severe
in view of the tight energy budget, as discussed above.

Identification of high energy neutrino signals from GRBs are facilitated through time coincidence,
but studying individual bursts in detail may be difficult \cite{bec06}.
Thus, photon signatures of UHECR production will also be important \cite{da06}.
We have conducted a detailed study of this issue for internal shocks
utilizing a comprehensive Monte Carlo code that includes a wide variety of relevant processes,
and found that in some cases, unique features such as synchrotron radiation
from muons and protons may be observable by GLAST and Cerenkov telescopes
(Asano \& Inoue, in preparation).

\section{Heavy nuclei from cluster accretion shocks as UHECRs}
\label{sec:clus}

In the currently favored picture of hierarchical structure formation in the CDM cosmology,
all massive clusters of galaxies should be surrounded by strong accretion shocks,
as a consequence of continuing infall of dark matter and baryonic gas \cite{min00}.
Such shocks should be interesting sites of particle acceleration,
and have been proposed as sources of UHECRs \cite{nma95, krj96}.
Here we briefly summarize our recent work on this subject invoking UHECR nuclei;
more details can be found in Ref. \refcite{isma07}.

For clusters of mass $M$, the rate of gas kinetic energy dissipation through accretion shocks
can be estimated as
$L_{\rm acc} \simeq 9 \times 10^{45} (M/{10^{15} M_\odot})^{5/3}$ erg/s \cite{kwl04}.
This can be combined with the Press-Schechter mass function
to evaluate $dE_{\rm kin}/dz$ for clusters of different $M$ as shown in Fig.\ref{fig:ene}.
Note that due to the hierarchical nature of structure formation
together with the nonlinear nature of gravity,
the maximum is reached at $z=0$,
with ample room to supply the UHECR energy budget.

\begin{figure}
\begin{center}
\psfig{file=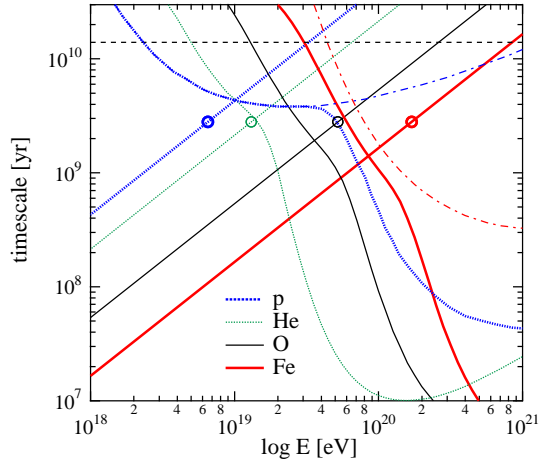,width=3.0in}
\end{center}
\caption{
Comparison of timescales at cluster accretion shocks
for shock acceleration $t_{\rm acc}$ (diagnonal lines),
and energy losses from interactions with background radiation fields (curves),
for protons (thick dotted), He (thin dotted), O (thin solid) and Fe nuclei (thick solid).
The photopair timescales are denoted separately for p and Fe (dot-dashed).
Also indicated are the Hubble time $t_H$ (dashed line)
and the escape-limited $E_{\max}$ (circles).
}
\label{fig:tacc}
\end{figure}

However, estimates of the maximum energy $E_{\max}$ for protons
seem to fall short of $10^{20}$ eV by 1-2 orders of magnitude \cite{krj96, ias05}.
A fiducial cluster of $M=2 \times 10^{15} M_\odot$
has shock radius $R_s \simeq 3.2$ Mpc
and shock velocity $V_s=(4/3)(GM/R_s)^{1/2} \simeq 2200$ km/s.
The shock magnetic field is taken to be $B_s=1 \mu$G, as suggested by some recent observations \cite{fn06}.
The timescale for shock acceleration
is $t_{\rm acc}=20 \kappa(E)/V_s^2 = (20/3) (E c/Z e B_s V_s^2)$,
assuming the Bohm limit for the diffusion coefficient  $\kappa(E)$
as inferred for supernova remnant shocks \cite{hil06}.
To be compared are the energy loss timescales
for photopair and photopion interactions with the CMB,
the escape time from the acceleration region $t_{\rm esc} \sim R_s^2/5 \kappa(E)$,
and the Hubble time $t_H$.
As is clear in Fig.\ref{fig:tacc}, for protons $E_{\max} \sim 10^{18}$-$10^{19}$ eV, confirming previous findings.

On the other hand, heavy nuclei with higher $Z$ have correspondingly shorter $t_{acc}$,
and Fe may be accelerated up to $10^{20}$ eV in the same conditions,
notwithstanding energy losses by photodisintegration with the FIRB and CMB (Fig.\ref{fig:tacc}).
In order to explore whether nuclei from cluster accretion shocks
can provide a viable picture of UHECR origin,
detailed propagation calculations of UHE nuclei above $10^{19}$ eV are undertaken,
following energy losses in the CMB/FIRB
and deflections in extragalactic magnetic fields (EGMF) for all particles
including secondary nuclei arising from photodisintegration.
We consider EGMF models that trace large-scale structure as in Ref. \refcite{asm05},
as well as the case of negligible EGMF, although Galactic fields are not included.
The source density is $n_s=2 \times 10^{-6} {\rm Mpc^{-3}}$, appropriate for massive clusters.
A fraction $f_{\rm CR}$ of the accretion luminosity $L_{\rm acc}$ is converted to cosmic rays
with energy distributions $\propto E^{-\alpha}$,
and we set $E_{\max}/Z=10^{19}$ eV, a fair approximation
to estimates for each species obtained 
by comparing timescales as in Fig.\ref{fig:tacc}.
For the elemental composition at injection, the He/p ratio is taken to be 0.042.
All heavier elements are assumed
to have the same relative abundances at fixed energy/nucleon
as that of Galactic cosmic ray sources at GeV energies \cite{all05},
and scaled with respect to protons by the metallicity $\zeta$ of the accreting gas.
We take $\zeta=0.2$ as suggested by both observations and theory \cite{nic05}.
An additional factor $A^{\beta}$ for the injected abundance of nuclei with mass number $A$
is introduced to take account of possible enhancement of heavier nuclei
due to nonlinear modification of shock structure by CRs \cite{bk99}.

\begin{figure}
\begin{center}
\psfig{file=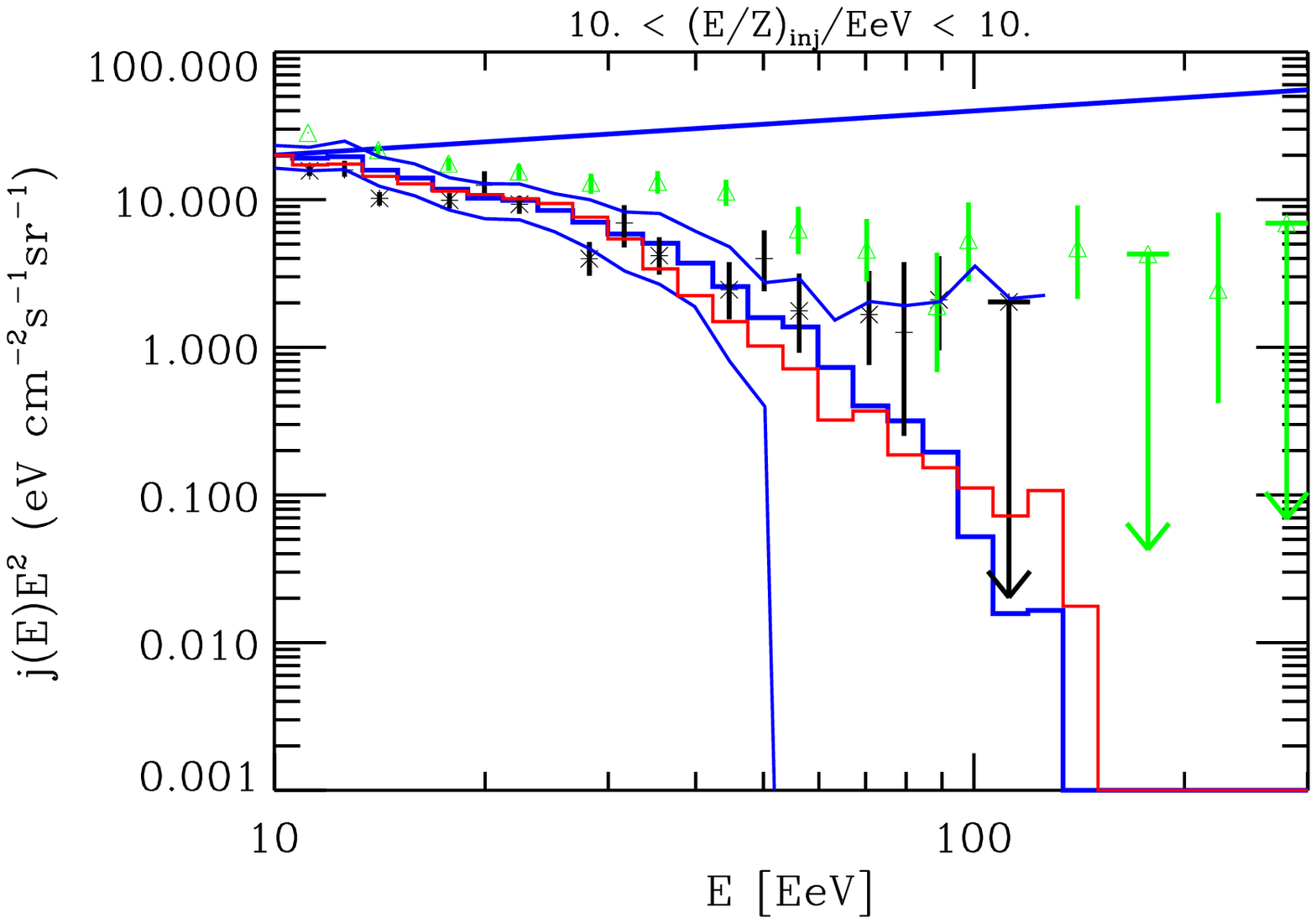,width=3in}
\psfig{file=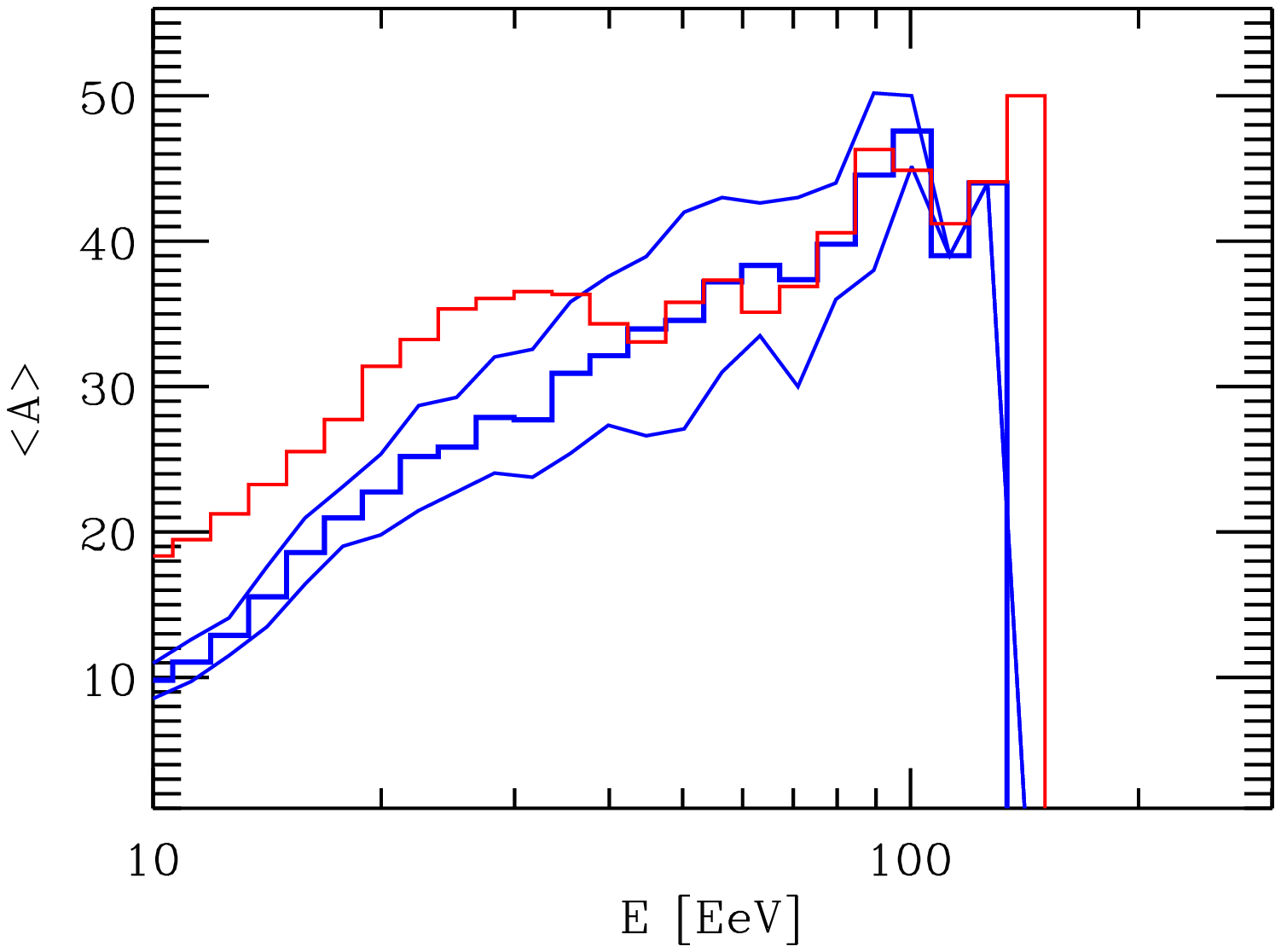,width=3in}
\end{center}
\caption{Observed UHECR spectrum (top) and mean mass composition (bottom)
versus energy $E$ (1 EeV $\equiv 10^{18}$ eV)
from cluster accretion shocks for $\alpha=1.7$ and $\beta=0.5$, compared with the current data
for AGASA \cite{tak98} (triangles), HiRes \cite{abb04} (bars) and Auger \cite{som05} (stars).
The histograms are the average result over different model realizations
for the cases with (thick) and without (thin) extragalactic magnetic fields,
and the thin curves outline the cosmic variance for the former case only.
The straight line in the top panel denotes $\alpha$.
See Ref. \refcite{isma07} for more details.}
\label{fig:speccomp}
\end{figure}

Fig.\ref{fig:speccomp} shows our results for the observed spectrum and composition
for $\alpha=1.7$ and $\beta=0.5$, which are quite consistent with the current data
for HiRes and Auger (and possibly AGASA as well \cite{isma07}).
Values of $\alpha<2$ are naturally expected at the high energy end
for nonlinear shock acceleration that accounts for the dynamical back reaction from CRs \cite{md01}.
The spectral steepening at $\gtrsim 10^{20}$ eV is due both to propagation losses
and the $E_{\max}$ limit at the source.
Normalization to the observed flux and comparison with the available accretion power
for $M > 10^{15} M_\odot$ fixes $f_{\rm CR}$,
which is $\simeq 0.01-0.6$ for cases with EGMF and $\simeq 0.004$ for the case without.
Low values of $f_{CR}$ may reflect inefficient escape of CRs from the system,
which is conceivable in view of the converging nature of the accretion flow.
CR escape may be mediated mainly during episodic merging events
that partially disrupt the shock and drive outflows of some of the downstream gas \cite{tak99}.

The mass composition at $\lesssim 3 \times 10^{19}$ eV is predominantly light
and consistent with HiRes claims \cite{abb05},
while the rapid increase of the average mass at higher energies
is a clear prediction of the scenario to be tested by the new generation experiments.
Despite the relative rarity of massive clusters in the local universe,
strong deflections of the highly charged nuclei in EGMF
allow consistency with the currently observed global isotropy.
On the other hand, with a sufficient number of accumulated events,
clear anistropies toward a small number of individual sources should appear,
although this prediction is subject to uncertainties in the EGMF and Galactic fields.
An aspect of this scenario that warrants further study is the spectral domain $< 10^{19}$ eV
and the implications for the Galactic-extragalactic transition region \cite{bgg06,all05}.

\begin{figure}
\begin{center}
\psfig{file=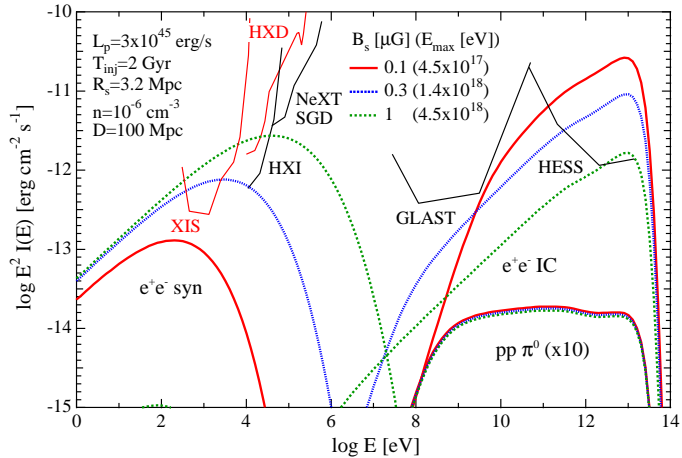,width=3.7in}
\end{center}
\caption{Spectra of UHE proton-induced photopair emission from the accretion shock
of a Coma-like cluster, for $B_s =$0.1, 0.3 and 1 $\mu$G.
The sensitivities for a 1 degree extended source are overlayed for HESS, GLAST,
Suzaku XIS+HXD, and NeXT HXI+SGD.}
\label{fig:uhexg}
\end{figure}

In this picture, some neutrino production may occur due to confined UHE protons
interacting with the enhanced FIRB toward the cluster center \cite{dem06},
along with a component from decay of photodisintegrated neutrons \cite{all06}.
Furthermore, we may look forward to very unique signatures in X-rays and gamma-rays.
Protons accelerated to $10^{18}$-$10^{19}$ eV in cluster accretion shocks
should efficiently channel energy into pairs of energy $10^{15}$-$10^{16}$ eV
through interactions with the CMB,
which then emit synchrotron radiation peaking in hard X-rays
and inverse Compton radiation in TeV gamma-rays.
Fig.\ref{fig:uhexg} displays the predicted spectra for a Coma-like cluster,
conservatively assuming that UHE proton injection continued only for a dynamical time $\simeq 2$ Gyr
(see Ref. \refcite{ias05} for more details).
The detection prospects are very promising
for Cerenkov telescopes such as HESS,
and hard X-ray observatories such as Suzaku and the future NeXT mission.
Photopair production by nuclei may also be efficient
and induce further interesting signals that are worth investigating (Fig.\ref{fig:tacc}).

Who is the real culprit flinging those nasty UHECRs at us?
AGNs, GRBs, clusters or some other guy?
Through the concerted effort
of upcoming cosmic ray, neutrino, X-ray, gamma-ray as well as radio observations,
we may finally find out who dunnit!

\section*{Acknowledgments}
The author thanks F. Aharonian, E. Armengaud, K. Asano, F. Miniati, G. Sigl and N. Sugiyama
for past and ongoing collaborations.


\begin{thebibliography}{9}
\bibitem{nw00} M. Nagano and A. A. Watson, {\em Rev. Mod. Phys.} {\bf 72}, 689 (2000);
   J. W. Cronin, {\em Nucl. Phys. B Proc. Suppl.} {\bf 138}, 465 (2005).
\bibitem{hil06} A. M. Hillas, astro-ph/0607109.
\bibitem{bs00} P. Bhattacharjee and G. Sigl, {\em Phys. Rep.} {\bf 327}, 109 (2000);
   R. J. Protheroe and R. W. Clay, {\em Pub. Astron. Soc. Aus.} {\bf 21}, 1 (2004).
\bibitem{gzk66} K. Greisen, {\em Phys. Rev. Lett.} {\bf 16}, 748 (1966);
   G. T. Zatsepin and V. A. Kuzmin, {\em JETP Lett.} {\bf 4}, 78 (1966).
\bibitem{psb76} J. L. Puget, F. W. Stecker and J. H. Bredekamp, {\em ApJ} {\bf 205}, 638 (1976).
\bibitem{ss99} F. W. Stecker and M. H. Salamon, {\em Astrophys. J.} {\bf 512}, 521 (1999);
   D. Hooper, S. Sarkar and A. M. Taylor, astro-ph/0608085.
\bibitem{abb05} R. U. Abbasi et al., {\em Astrophys. J.} {\bf 622}, 910 (2005).
\bibitem{ave03} M. Ave et al., {\em Astropart. Phys.} {\bf 19}, 61 (2003).
\bibitem{wat04} A. A. Watson, {\em Nuc. Phys. B Proc. Suppl.} {\bf 136}, 290 (2004).
\bibitem{kro94} P. P. Kronberg, {\em Rep. Prog. Phys.} {\bf 57}, 325 (1994).
\bibitem{sme03} G. Sigl, F. Miniati and T. A. Ensslin,
   {\em Phys. Rev.} D {\bf 70}, 43007 (2004).
\bibitem{dol04} K. Dolag et al.,
   {\em JCAP} {\bf 1}, 9 (2005).
\bibitem{tak06} H. Takami, H. Yoshiguchi and K. Sato, {\em Astrophys. J.} {\bf 639}, 803 (2006);
   M. Kachelrie\ss, P. D. Serpico and M. Teshima, astro-ph/0510444.
\bibitem{gae06} B. M. Gaensler, {\em Astron. Nachr.} {\bf 327}, 387 (2006).
\bibitem{hil84} A. M. Hillas, {\em Ann. Rev. Astron. Astrophys.} {\bf 22}, 425 (1984);
  D. Torres and L. A. Anchordoqui, {\em Rep. Prog. Phys.} {bf 67}, 1663 (2004). 
\bibitem{is01} S. Inoue and S. Sasaki, {\em Astrophys. J.} {\bf 562}, 618 (2001).
\bibitem{pm01} C. Porciani and P. Madau, {\em Astrophys. J.} {\bf 548}, 522 (2001).
\bibitem{bgg06} V. Berezinsky, A. Gazizov and S. Grigorieva, {\em Phys. Rev.} D {\bf 74}, 43005 (2006).
\bibitem{mpr01} K. Mannheim, R. J. Protheroe and J. P. Rachen, {\em Phys. Rev.} D {\bf 63}, 3003 (2001).
\bibitem{bs87} P. L. Biermann and P. A. Strittmatter, {\em Astrophys. J.} {\bf 322}, 643 (1987);
  J. P. Rachen and P. L. Biermann, {\em Astron. Astrophys.} {\bf 272}, 161 (1993).
\bibitem{nma95} C. A. Norman, D. B. Melrose and A. Achterberg, {\em Astrophys. J.} {\bf 454}, 60 (1995).
\bibitem{hh02} F. Halzen and D. Hooper, {\em Rep. Prog. Phys.} {\bf 65}, 1025 (2002).
\bibitem{ach05} A. Achterberg et al., {\em Astropart. Phys.} {\bf 26}, 282 (2005).
\bibitem{aha02} F. A. Aharonian, {\em Mon. Not. R. A. S.} {\bf 332}, 215 (2002).
\bibitem{ga05} S. Gabici and F. A. Aharonian, {\em Phys. Rev. Lett.} {\bf 95}, 251102;
  E. Armengaud, G. Sigl and F. Miniati, {\em Phys. Rev.} D {\bf 73}, 083008 (2006).
\bibitem{wax95} E. Waxman, {\em Phys. Rev. Lett.} {\bf 75}, 386 (1995);
  M. Vietri, {\em Astrophys. J.} {\bf 453}, 883 (1995).
\bibitem{vdg03} M. Vietri, D. De Marco and D. Guetta, {\em Astrophys. J.} {\bf 592}, 378 (2003);
  E. Waxman, {\em Astrophys. J.} {\bf 606}, 988 (2004).
\bibitem{asa05} K. Asano, {\em Astrophys. J.} {\bf 623}, 967 (2005).
\bibitem{bec06} J. Becker et al. {\em Astropart. Phys.} {\bf 25}, 118 (2006).
\bibitem{da06} C. D. Dermer and A. Atoyan, {\em New J. of Phys.} {\bf 8}, 122 (2006).
\bibitem{isma07} S. Inoue, G. Sigl, F. Miniati and E. Armengaud, {\em Phys. Rev. Lett.}, submitted (astro-ph/0701167).
\bibitem{min00} F. Miniati et al., {\rm Astrophys. J.} {\bf 542}, 608 (2000);
  D. Ryu et al., {\em Astrophys. J.} {\bf 593}, 599 (2003).
\bibitem{krj96} H. Kang, D. Ryu and T. W. Jones, {\em Astrophys. J.} {\bf 456}, 422 (1996);
  H. Kang, J. P. Rachen, P. L. Biermann, {\em Mon. Not. R. A. S.} {\bf 286}, 257 (1997);
  M. Ostrowski and G. Siemieniec-Ozi\c{e}b\l o, {\em Astron. Astrophys.} {\bf 386}, 829 (2002).
\bibitem{kwl04} U. Keshet, E. Waxman and A. Loeb, {\em Astrophys. J.} {\bf 617}, 281 (2004);
  V. Pavlidou and B. D. Fields,  {\em Astrophys. J.} {\bf 642}, 734 (2006).
\bibitem{ias05} S. Inoue, F. A. Aharonian and N. Sugiyama, {\em Astrophys. J.} {\bf 628}, L9 (2005).
\bibitem{fn06} L. Feretti and D. M. Neumann, {\em Astron. Astrophys.} {\bf 450}, L21 (2006);
  M. Johnston-Hollitt and R. Ekers, astro-ph/0411045.
\bibitem{asm05} E. Armengaud, G. Sigl and F. Miniati, {\em Phys. Rev.} D {\bf 72}, 043009 (2005).
\bibitem{all05} D. Allard et al., {\em Astron. Astrophys.} {\bf 443}, L29 (2005);
  G. Sigl and E. Armengaud, {\em JCAP} {\bf 0510}, 016 (2005).
\bibitem{nic05} F. Nicastro et al. {\em Nature} {\bf 433}, 495 (2005);
  R. Cen and J. P. Ostriker, astro-ph/0601008.
\bibitem{bk99} E. G. Berezhko and L. T. Ksenofontov, {\em JETP} {\bf 89}, 391 (1999);
  L. O'C. Drury et al., {\em Sp. Sci. Rev.} {\bf 99}, 329 (2001).
\bibitem{tak98} M. Takeda et al., {\em Phys. Rev. Lett.} {\bf 81}, 1163 (1998).
\bibitem{abb04} R. U. Abbasi et al., {\em Phys. Rev. Lett.} {\bf 92}, 151101 (2004).
\bibitem{som05} P. Sommers [Pierre Auger collaboration], 29th ICRC\ {\bf 7}, 387 (2005).
\bibitem{md01} M. A. Malkov and L. O'C. Drury, {\em Rep. Prog. Phys.} {\bf 64}, 429 (2001);
  H. Kang and T. W. Jones, {\em Astrophys. J.} {\bf 620}, 44 (2005).
\bibitem{tak99} M. Takizawa, {\em Astrophys. J.} {\bf 520}, 514 (1999).
\bibitem{dem06} D. De Marco et al., {\em Phys. Rev.} D {\bf 73}, 043004 (2006).
\bibitem{all06} D. Allard et al., {\em JCAP} {\bf 0609}, 005 (2006).
\end{thebibliography}
\end{document}